\documentclass[usenatbib]{mn2e}
\usepackage{epsfig}
\usepackage{subfigure}
\usepackage{amssymb}
\usepackage{color}

\newcommand{\bn}{\begin{enumerate}}
\newcommand{\en}{\end{enumerate}}
\newcommand{\bi}{\begin{itemize}}
\newcommand{\ei}{\end{itemize}}

\newcommand{\Msun}{M_{\odot}}
\newcommand{\Mstar}{M_{\star}}
\newcommand{\Zsun}{Z_{\odot}}
\newcommand{\Mhalo}{M_{\rm halo}}
\newcommand{\vwind}{v_{\rm wind}}
\newcommand{\vesc}{v_{\rm esc}}
\newcommand{\siggal}{\sigma_{\rm gal}}
\newcommand{\rhoth}{\rho_{\rm th}}
\newcommand{\Civ}{C\,{\sc iv}}

\newcommand{\apj}{ApJ}

\newcommand{\apjl}{ApJL}
\newcommand{\apjs}{ApJS}
\newcommand{\mnras}{MNRAS}
\newcommand{\aj}{AJ}
\newcommand{\araa}{ARA\&A}

\newcommand{\gca}{GeCoA}
\newcommand{\pasp}{PASP}

\title{Multicomponent and Variable Velocity Galactic Outflow in Cosmological Hydrodynamic Simulations}
\author[Choi \& Nagamine]
{Jun-Hwan Choi\thanks{Email: jhchoi@physics.unlv.edu} Kentaro Nagamine\thanks{Visiting Researcher, Institute for the Physics and Mathematics 
of the Universe (IPMU), University of Tokyo, 5-1-5 Kashiwanoha, 
Kashiwa, 277-8583, Japan} \vspace{0.2cm}\\
Department of Physics and Astronomy, University of Nevada, Las Vegas,  
4505 S. Maryland Pkwy, Las Vegas, NV, 89154-4002, U.S.A.}

\begin{document}

\maketitle

\begin{abstract}
We develop a new ``Multicomponent and Variable Velocity'' (MVV) galactic outflow model for cosmological smoothed particle hydrodynamic (SPH) simulations.
The MVV wind model reflects the fact that the wind material can arise from different phases in the interstellar medium (ISM), and the mass-loading factor in the MVV model is a function of galaxy stellar mass. 
We find that the simulation with the MVV outflow has the following characteristics:
(i) the intergalactic medium (IGM) is hardly heated up, and the mean IGM temperature is almost the same as in the no-wind run;
(ii) it has lower cosmic star formation rates (SFRs) compared to the no-wind run, but higher SFRs than the constant velocity wind run;
(iii) it roughly agrees with the observed IGM metallicity, and roughly follows the observed evolution of $\Omega$(\Civ);
(iv) the lower mass galaxies have larger mass-loading factors, and the low-mass end of galaxy stellar mass function is flatter than in the previous simulations.
Therefore, the MVV outflow model mildly alleviates the problem of too steep galaxy stellar mass function seen in the previous SPH simulations.
In summary, the new MVV outflow model shows reasonable agreement with observations, and gives better results than the constant velocity wind model.
\end{abstract}



\section{Introduction}
\label{sec:intro}

It has been pointed out by many researchers that galactic outflow plays a crucial role in galaxy formation.
For example, the physical size and mass of a disk galaxy might be determined by the balance between gas accretion and outflow \citep{Dekel.Silk:86,White.Frenk:91}.
Galactic outflows also regulate the star formation by removing gas from galaxies, and help to alleviate the ``overcooling'' problem seen in the earlier simulations of galaxy formation.
In addition, the observations of quasar absorption lines reveal that the intergalactic medium (IGM) is considerably enriched with metals, and galactic outflow is thought to be the responsible mechanism to spread metals into the IGM \citep[e.g.,][]{Aguirre.etal:01, Cen.etal:05}.
Therefore, an appropriate model of galactic outflow has to be implemented in all cosmological simulations of galaxy formation in order to reproduce the observed statistics of galaxies and IGM.

There are two major astrophysical sources that drive the galactic outflow: supernovae (SNe) and supermassive black holes (SMBHs).
The SN feedback probably plays an important role throughout the entire history of the universe as long as the star formation is ongoing, and it is considered to be responsible for enriching the IGM.
The SMBH feedback is considered to play an important role in determining the evolution of massive galaxies at low-$z$ \citep[e.g.,][]{Keres.etal:09}.
In this paper, we focus on the outflows driven by SNe, and examine the effects of different models of galactic outflows on galaxy growth and IGM enrichment.

Recent observations show that the galactic outflow consists of multiphase gas.
Soft X-ray observations reveal that some galaxies are surrounded by haloes of hot gas \citep{Strickland.Stevens:00}.
A diffuse hard X-ray emission that would be associated with hot plasma of temperature $T=10^7 - 10^8$\,K was detected by \emph{Chandra X-ray Observatory} in M82 \citep{Strickland.Heckman:09}.
\citet{Martin:05,Martin:06} studied the Na\,{\sc i} absorption lines in ultra-luminous infrared galaxies, and argued that a significant fraction of galactic outflow contains relatively cool (or warm neutral/ionised) gas.
Although there are many ongoing observational studies, the detailed composition of multiphase outflow is still poorly constrained.

The absorption line studies of outflowing gas from starburst galaxies suggest that the outflow velocity is a function of galaxy mass or star formation rate (SFR) \citep{Martin:05,Rupke.etal:05}.
Using a large sample of low-resolution spectroscopy for $z \sim 1$ galaxies from the DEEP2 survey, \citet{Weiner.etal:09} argued that the outflow velocities of $z \sim 1$ galaxies scale as $\vwind \propto {\rm SFR}^{0.3}$.
\citet{Martin:05} also reported that $\vwind \propto {\rm SFR}^{0.35}$
These observations attempt to make connections with theoretical models of wind driving mechanisms \citep[e.g.,][]{Murray.etal:05}, but a clear connection between the theory and observations is yet to be established.

Our current understanding of the wind driving mechanisms and the propagation processes of outflow from first principles is still crude.
Although there have been some recent progress using high-resolution hydrodynamic simulations in generating galactic outflows and resolving the evolution of shocked, swept-up shell \citep{MacLow.Ferrara:99,Mori.etal:02,Fujita.etal:09}, many uncertainties remain regarding the energy source of the wind, instabilities in the outflowing medium, and the interaction between the wind and ambient gas.

Moreover, in cosmological hydrodynamic simulations, there is a so-called ``overcooling'' problem: if the SN feedback energy is deposited as a thermal energy and the simulation doesn't have an adequate resolution to resolve the expanding hot bubble of gas, the thermal energy is quickly radiated away without having any effect on the surrounding medium \citep[e.g.,][]{Katz.etal:96,Ceverino.Klypin:09}.
In the real Universe, overlapping SN explosions form a low-density superbubble of hot gas, which lasts longer and significantly affects the evolution of surrounding medium by suppressing subsequent cooling of gas. 
These hot superbubbles cannot be resolved even in the highest resolution cosmological simulation of today's, if one wants to follow the dynamical evolution of numerous galaxies simultaneously in a comoving volume of $>$10\,Mpc.  
Therefore, the present-day cosmological hydro simulations could still suffer from the overcooling problem if no further treatment is given. 

To avoid this problem, many cosmological simulations adopt a phenomenological model of galactic outflow.
One widely adopted treatment employs a subgrid multiphase ISM model and the ejection of SPH particles as winds \citep{Springel.Hernquist:03,Tornatore.etal:04,Oppenheimer.Dave:06}.
In this model, the SN energy is deposited in the hot ISM, and the galactic outflow is generated by temporally converting some gas particles into wind particles based on the mass-loading factor: 
\begin{eqnarray}
\eta \equiv \dot{M_w}/\dot{\Mstar},
\end{eqnarray}
where $\dot{M_w}$ and $\dot{\Mstar}$ are the wind mass transfer rate and galaxy SFR, respectively.  
The wind particles get the initial kick according to the amount of SN energy injection, and all wind particles initially have the same velocity and the same $\eta$, i.e., the ``constant velocity wind'' model.
During the wind phase, the particles are temporally decoupled from the hydro force and are able to escape from the galaxy.
This wind particle method can enrich the IGM, and regulate the galaxy growth and star formation. 

We point out that there are several other attempts to model galaxy outflows in cosmological simulations.
\citet{Stinson.etal:06} proposed to suppress gas cooling during the expansion of SN bubble phase, whose size is computed by a blast-wave model.
This approach allows them to avoid the ``overcooling'' problem.
\citet{ScannapiecoC.etal:06} proposed to explicitly divide the SPH particles into cold and hot phase particles, and performed the SPH smoothing separately for each phases to avoid the artificial cooling caused by the mixing of different phases. 
Their model also suppresses the cooling of SN energy injected into the cold phase, and converts the cold phase to the hot phase after a certain amount of SN energy injection.
\citet{DallaVecchia.Schaye:08} implemented the kinetic feedback by injecting the SN kinetic energy into the neighbouring gas particles. 
Each of these attempts have remedied the overcooling problem to some extent and improved the galactic outflow model in cosmological hydrodynamic simulations.

Although the wind particle method developed by \citet{Springel.Hernquist:03} and other alternative approaches have been successful to some level, there are still remaining issues.
First, there are two theoretical ideas for the wind driving mechanism --- the ``energy-driven'' wind and the ``momentum-driven'' wind model.
In the energy-driven model, the thermal energy injected by SNe into the ambient medium drives the galactic wind through the expansion of hot bubbles. 
In the momentum-driven model, the momentum applied by the radiation pressure from massive stars pushes the dust particles, which entrain the gas to drive galactic winds \citep{Murray.etal:05}.
An important point is that the two models have different values of $\eta$ parameter: 
\begin{eqnarray}
\eta \propto
\left \{
\begin{array}{ll}
\siggal^{-2} & {\rm (energy-driven)}\\
\siggal^{-1}  & {\rm (momentum-driven)},
\end{array}
\right.
\end{eqnarray}
where $\siggal$ is the velocity dispersion of the galaxy \citep{Murray.etal:05}.  
However, we note that these relationships are derived from simple scaling arguments and may not be so robust.  
The values of $\eta$ are only loosely constrained by observations \citep[e.g.,][]{Martin:99}. 
The relative importance of the two theoretical ideas has not been considered carefully in cosmological hydrodynamic simulations. 

Second, there are many problems in the previously adopted wind models.
For example, the constant velocity wind model seems to predict too little C{\sc iv} absorption, too small C{\sc iii}/C{\sc iv} ratios \citep{Aguirre.etal:05}, and overheat the IGM \citep{Oppenheimer.Dave:06}.  
The latter work also argued that the momentum-driven wind model produces more mass-loaded but low velocity winds from early galaxies, so as to eject a substantial metal mass into the IGM without over-heating it.

Third, the multiphase nature of the wind medium has not been recognised very much for the galactic wind models in cosmological simulations.
In the earlier work, the wind particles were selected only from star-forming particles or their neighbours.
Owing to the hot gas heated by the SNe in the subgrid ISM model, the gas temperature in and around the star-forming region is very hot.
In addition, the high constant wind velocity may have heated the IGM too much, and we need to revise the wind model to incorporate a cold component into the wind medium. 

The best way to resolve these issues is to perform hydrodynamic simulations with a very high-resolution, and simulate the galactic outflow in a cosmological environment from first principles.
However, the computational expense of such a cosmological simulation is still very high and not possible at this moment.
An alternative approach is to develop more realistic phenomenological galactic outflow models, verify the model by comparing with observations, and we can gain deeper insight into the phenomenon of galactic outflow.
In this paper, we adopt the latter approach to study the role of galactic outflow in galaxy formation, and develop a new phenomenological wind model.
The new outflow model relies on the observational constraints: the wind velocity ($\vwind$) as a function of galaxy SFR, and $\eta$ based on galaxy mass. 
Using this phenomenological approach, the new model captures the multicomponent origin of wind material and variable wind velocity.

This paper is organized as follows. 
In Section~\ref{sec:method}, we describe our simulation with a focus on the new wind model.
In Section~\ref{sec:global}, we present some of the global statistics such as the overall distribution of gas temperature and metals, cosmic SFR, and galaxy mass functions.
In Section~\ref{sec:IGM}, we focus on the IGM properties such as the mass fractions in various phases, overdensity--metallicity relationship, and the evolution of cosmic C{\sc iv} mass density.
In Section~\ref{sec:summary}, we will summarise our findings and discuss the effects of our new wind model.

\section{Numerical technique}
\label{sec:method}

\subsection{Simulation Setup}

We will use the modified version of the Tree-particle-mesh (TreePM) smoothed particle hydrodynamics (SPH) code GADGET-3 \citep[originally described in][]{Springel:05}.
In this code, the SPH calculation is performed based on the entropy conservative formulation \citep{Springel.Hernquist:02}.
Our conventional code includes radiative cooling by H, He, and metals \citep{Choi.Nagamine:09a}, heating by a uniform UV background of a modified \citet{Haardt.Madau:96} spectrum \citep{Katz.etal:96,Dave.etal:99}, star formation, supernova feedback, phenomenological model for galactic winds, and a sub-resolution model of multiphase ISM \citep{Springel.Hernquist:03}.

In the multiphase ISM model, high-density ISM is pictured to be a two-phase fluid consisting of cold clouds in pressure equilibrium with a hot ambient phase.
Cold clouds grow by radiative cooling out of the hot medium, and this material forms the reservoir of baryons available for star formation.

For the star formation model, we adopt the ``Pressure model'' described in \citet{Choi.Nagamine:09b}.
This star formation model estimates the SFR based on the gas pressure rather than gas density, and implicitly considers the effect of H$_2$ formation on star formation \citep{Schaye.DallaVecchia:08,Choi.Nagamine:09b}.
Interested readers can refer to \citet{Choi.Nagamine:09a,Choi.Nagamine:09b} for detailed descriptions.

\subsection{Multicomponent Variable Velocity Wind Model}
\label{sec:MVV}

We will call our phenomenological model of galactic wind as the ``Multicomponent Variable Velocity'' (MVV) wind, because the wind material is taken from both low temperature, high density regions and high temperature, low density regions. 
The fundamental parameters of our MVV model model are $\eta$ and $\vwind$, which are constrained by the observations. 
As mentioned in Section~\ref{sec:intro}, there are numerical limitations as to properly reproduce the formation of galactic outflow.
In addition, newly proposed radiation pressure model is based on a significantly different wind launching mechanism compared to the well-developed supernova-driven wind.
Therefore, we use observational constraints to compute $\eta$ and $\vwind$.

Most observations provide the values of $\eta$ and $\vwind$ as a function of host galaxy mass/SFR. 
To compute the galaxy mass and SFR for galactic winds, we have implemented a group-finder algorithm into GADGET-3 to compute the galaxy masses and SFR on-the-fly while the simulation is running.
The group-finder, which is a simplified variant of the \textup{SUBFIND} algorithm developed by \citet{Springel.etal:01}, identifies the isolated groups of star and gas particles (i.e., galaxies) based on the baryonic density field.  
The detailed procedure of this galaxy grouping is described in \citet{Nagamine.etal:04}. 
The outer baryonic density threshold for a galaxy is 0.01$\rhoth$, where $\rhoth$ is the threshold density for star formation. 
The group properties such as the stellar mass, SFR, and maximum size of the galaxy are distributed to and stored by the individual member particles.
During the course of simulation, the group-finder is invoked whenever the total SFR or total stellar mass increases by 10\% from the previous group finding.
In addition, the group finding is also performed when the snapshot file is generated.
In particular, the frequency of group finding ($\Delta z$) is approximately 0.11 for the `1.5ME' run in the redshift range $3 < z < 5$.
The wind parameters ($\eta$ and $\vwind$) are computed based on the group properties stored in the member particle data, as we describe below.

In our MVV wind model, the chance that a gas particle becomes a wind particle is based on the values of $\eta$, galaxy mass, and galaxy SFR.
The model assumes that all gas particles in a given galaxy have the same chance to become part of the wind, which has an important consequence.
Since any gas particle inside a galaxy, either hot or cold, could become a wind particle, the galactic outflow naturally reproduces a multiphase nature of outflow medium.
The multiphase nature of galactic outflow is suggested by many observations \citep{Martin:05, Martin:06, Strickland.Heckman:09}, but it has not been taken into account explicitly in cosmological hydrodynamic simulations.
In our model, the wind particles are randomly chosen from constituent gas particles of a galaxy, because we currently have very limited knowledge on the origin of the multiphase outflow medium.
The outflow in our model includes gas from high density, cold gas in the galaxy, resembling the cold component entrained by the hot expanding gas from SNe \citep[e.g.,][]{Heckman:02}.
Note that the `multiphase' nature of the MVV wind model does not mean that a single wind particle is made of two or three phase medium, but it simply means that the new wind model generates a wind made of both hot and cold gas. 

\begin{figure*}
\centerline{
  \includegraphics[width=0.9\textwidth] {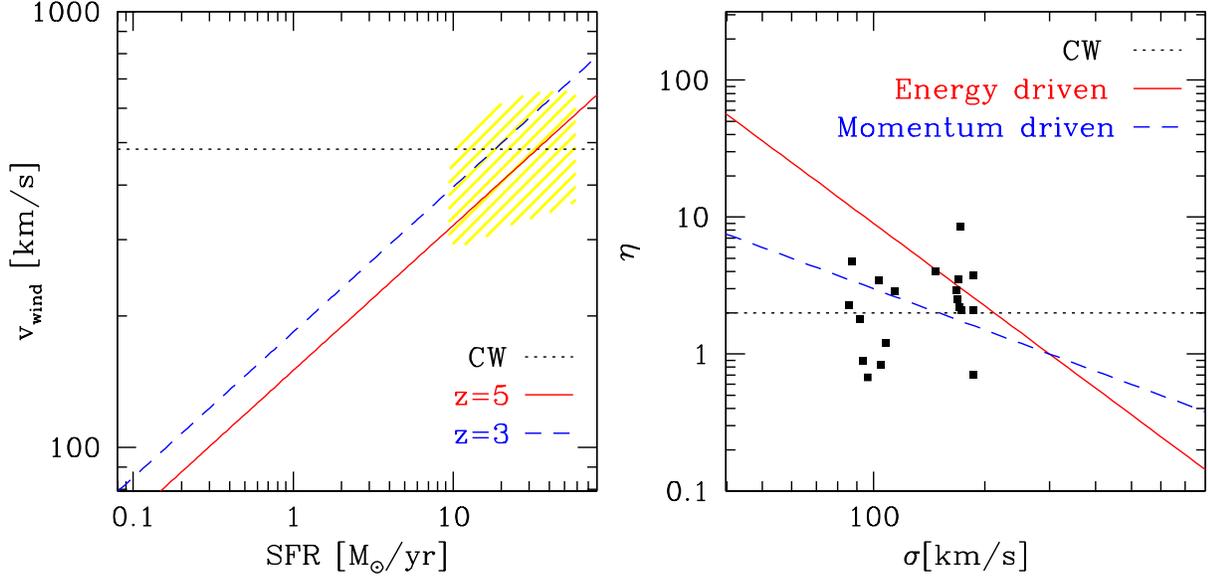}
}
\caption{
Input wind parameters ($\vwind$ and $\eta$) for the MVV wind model.
{\it Left:} Wind velocity as a function of galaxy SFR assuming $\zeta = 1$. 
We compare the $\vwind$ at $z = $3 and 5, as well as the constant wind velocity (484\,km\,s$^{-1}$) in the CW run.
The yellow shade is the observed wind velocity range from \citet{Weiner.etal:09}.
{\it Right:} Wind mass-loading factor as a function of galaxy velocity dispersion.
We compare $\eta$ for the energy driven wind and momentum driven wind, as well as the constant $\eta$ in the CW run.
The points are the observed data from \citet{Martin:99}.
}
\label{fig:windmodel}
\end{figure*}

In the MVV wind model, $\vwind$ depends on the galaxy mass and galaxy SFR.
We first assume that the wind velocity is proportional to the escape velocity of the galaxy: 
\begin{eqnarray}
\vwind = \zeta\, \vesc, 
\label{eq:vwind}
\end{eqnarray}
where $\zeta$ is a scaling factor of wind velocity.  
We then employ an empirical relationship between $\vesc$ and galaxy SFR as follows. 
From the halo mass ($\Mhalo$), we can compute the circular velocity as
\begin{eqnarray}
v_{c} & = & \left ( \frac{G\Mhalo}{R_{200}} \right )^{1/2} \; =  \; 
\left [ G \Mhalo^{2/3} \left (\frac{4 \pi}{3} \bar{\rho}_{200} \right )^{1/3} \right ]^{1/2} 
\nonumber \\
  & = & 124 \left (\frac{\Mhalo}{10^{11}\, h^{-1}\Msun} \right )^{1/3} \left (\frac{1+z}{4} \right )^{1/2} {\rm km\, s^{-1}},
  \label{eq:vc}
\end{eqnarray}
where $R_{200}$ and $\bar{\rho}_{200}$ is the radius and mean density of the halo with an overdensity of 200.

The SFR can be estimated from the UV luminosity: $SFR\, [\Msun\,{\rm yr}^{-1}] = 1.4 \times 10^{-28} L_{\nu}\, [{\rm erg\, s}^{-1}\, {\rm Hz}^{-1}]$ \citep{Kennicutt:98}.
Earlier, using the results from cosmological simulations, \citet{Nagamine.etal:07} obtained the relationship between the absolute AB magnitude of a galaxy and halo mass: $M_{AB} = -2.5(\log \Mhalo -12) -21.5$, where $\Mhalo$ is in units of $h^{-1}\Msun$. 
Combining the above two relationships, we obtain 
\begin{eqnarray}
\log (SFR) = \log (\Mhalo) -10.62.
\label{eq:SFR_M}
\end{eqnarray}
This relationship provides a crude, but a reasonable conversion between $\Mhalo$ and SFR.

Combining Equations (\ref{eq:vc}) and (\ref{eq:SFR_M}), and using $\vesc = \sqrt2 v_c$, we obtain the relationship between $\vesc$ and SFR as
\begin{eqnarray}
\vesc = 130 \, (SFR)^{1/3} \left (\frac{1+z}{4} \right )^{1/2} {\rm km\,s^{-1}}.
  \label{eq:SFR_vesc}
\end{eqnarray}
The above relationship  $\vesc \propto (SFR)^{1/3}$ is consistent with the observed one \citep{Martin:05,Weiner.etal:09}. 

As discussed in Section~\ref{sec:intro}, the two wind-driving mechanisms suggest two different scalings for $\eta$.
In the energy-driven wind case,
\begin{eqnarray}
\eta = (\sigma_{0}/\siggal)^2,
\end{eqnarray}
where $\sigma_{0}=300$\,km s$^{-1}$, and $\siggal = \vesc / 2$ is the velocity dispersion of the galaxy.
In the momentum-driven wind case, 
\begin{eqnarray}
\eta = \sigma_{0}/\siggal.
\end{eqnarray}
In Figure~\ref{fig:windmodel}, we show the scaling of $\eta$ and $\vwind$ as a function of galaxy SFR and velocity dispersion $\sigma$.  
In the left panel, one can see that, in the MVV model the wind velocity is an increasing function of SFR and redshift, in contrast to the constant wind velocity model. 
Unfortunately, it has not been understood which wind-driving mechanism is the dominant process, and the current observational data is not yet robust enough to clearly discern the best wind model.
Therefore, we develop a new wind model by employing both mechanisms in this paper.
In the low-density regions ($\rho_{\rm gas} < \rhoth$), the gas is sufficiently far away from star-forming regions, and the hot bubbles of gas are easy to expand and escape the region. 
Therefore in our fiducial wind model, we employ the energy-driven mechanism for the gas particles with $\rho_{\rm gas} < \rhoth$.
On the other hand, in the high-density star-forming regions ($\rho_{\rm gas} > \rhoth$), gas is close to the massive stars and SNe, and more likely to be subject to a strong radiation from these sources.  
Therefore in our fiducial wind model, we employ the momentum-driven mechanism for the gas particles with $\rho_{\rm gas} > \rhoth$.

Since the new model generates the wind particle from both star-forming and non-starforming regions, the $\eta$ parameter can be written as an addition of two 
separate parts:
\begin{equation}
\eta = \eta_{SF} + \eta_{non-SF}, 
\end{equation}
where 
$\eta_{SF} = \dot{M}_{w, SF}/\dot{\Mstar}$ for the wind from star-forming gas, and  $\eta_{non-SF} = \dot{M}_{w, non-SF}/\dot{\Mstar}$ for the wind from non-star-forming gas.  Therefore, 
\begin{equation}
\dot{M}_w = \dot{M}_{w, SF} + \dot{M}_{w, non-SF} = (\eta_{SF} + \eta_{non-SF}) \dot{\Mstar}
\end{equation}
for the total wind. 
Since the wind particles are selected randomly within a given galaxy, the mass fraction of the wind material that comes from star-forming and non-starforming gas would depend on the masses of the two components. 

We note that our new wind model is an alternative phenomenological model for galactic outflow in cosmological SPH simulations.
The values of wind parameters are determined by observational constraints rather than from physical first principles.
In addition, the wind particles are decoupled from the hydro force during the wind phase.
This assumption has been criticised by a few authors \citep[e.g.,][]{DallaVecchia.Schaye:08} owing to its unphysical nature. 
However, it is expected that the hydrodynamical interaction between wind particles and other gas particles in the galaxy could give rise to unwanted effects for low-resolution simulations.  
If galactic outflow cannot be launched due to hydrodynamic interaction, there is no point in implementing this phenomenological model in our simulation. 
In the future, we need to continuously improve our model so that the wind generation and propagation will be based on physical principles.

In the default model, we adopt $\zeta$=1.5 for the momentum-driven wind (for the high-density gas source) and $\zeta$=1 for the energy-driven wind (for the low-density gas source).  
We call this default model the ``1.5ME'' run. 
The value of $\zeta$ for the momentum-driven wind can be greater than unity, because the radiation pressure applied by the massive stars can continuously push the outflowing gas to a velocity greater than the escape velocity.
Besides our fiducial wind mixture, we also test several other mixtures of $\eta$ and $\zeta$, as shown in Table~\ref{table:model}
For comparison purpose, we also consider the models with no wind (NW) and a constant velocity wind (CW) \citep{Springel.Hernquist:03}.
In the CW run, only star-forming SPH particles can become wind particles, and fixed values of $\eta = 2$ and $\vwind = 484$\,km\,s$^{-1}$ are adopted.

\begin{table}
\begin{center}
\begin{tabular}{ccccc}
\hline
       &  \multicolumn{2}{c}{High density}  & \multicolumn{2}{c}{Low density} \\ 
Model  & $\eta$  &  $\zeta$ & $\eta$  &  $\zeta$  \\
\hline
\hline
1.5ME   & Momentum & 1.5  & Energy   & 1 \cr
ME      & Momentum & 1    & Energy   & 1 \cr
EE      & Energy   & 1    & Energy   & 1 \cr
1.5MM   & Momentum & 1.5  & Momentum & 1.5 \cr
NW      &   \multicolumn{4}{c}{No galactic wind}    \cr
CW      &   \multicolumn{4}{c}{Constant velocity wind} \cr
\hline
\end{tabular}
\caption{
The six different wind models employed in this paper.
The parameter $\eta$ is the mass-loading factor, and $\zeta$ is the scaling parameter for the wind velocity. 
As for the model name, for example, ``1.5ME'' consists of the value of $\zeta$, ``M(omentum)'', and ``E(nergy)'', for high and low density gas, respectively.   
}
\label{table:model}
\end{center}
\end{table}

In this paper, we adopt the latest WMAP5 cosmology: $\Omega_m = 0.26$, $\Omega_{\Lambda} = 0.74$,  $\Omega_b = 0.044$, $h=0.72$, $n_{s}=0.96$, and $\sigma_{8}=0.80$ \citep{Komatsu.etal:09}.
The simulations are initialised with 216$^3$ particles each for gas and dark matter in a comoving box of $(10\,h^{-1}$\,Mpc)$^3$.
We stop this series at $z=2.75$, as it misses the long wavelength perturbations at lower redshifts.
In these simulations, dark matter particle mass is $5.96 \times 10^6 h^{-1}\Msun$, and the initial gas particle mass is $1.21 \times 10^6 h^{-1} \Msun$.
Using the identical initial condition for all the runs, we examine the effects of different wind models listed in Table~\ref{table:model}.

\section{Global Properties}
\label{sec:global}

\subsection{Distribution of Gas, Thermal Energy, and Metals}

\begin{figure*}
\centerline{
  \includegraphics[width=0.3\textwidth,angle=-90] {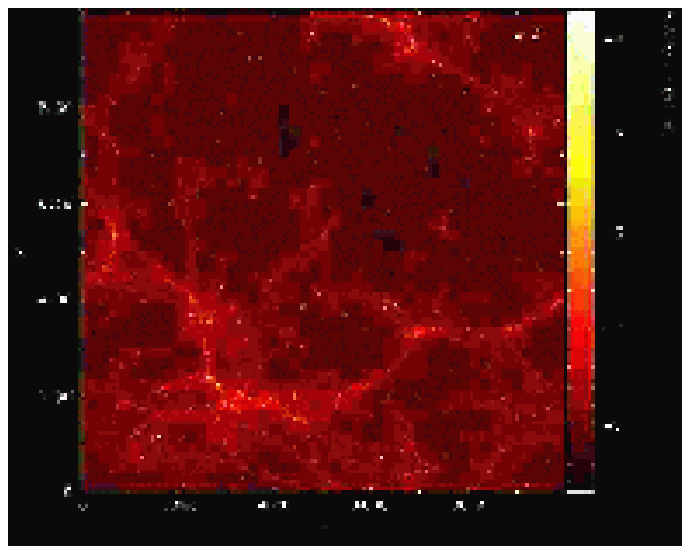}
  \includegraphics[width=0.3\textwidth,angle=-90] {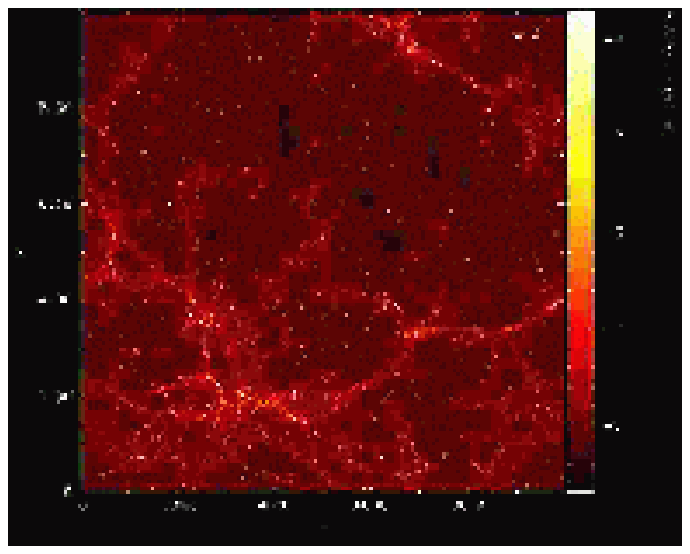}
  \includegraphics[width=0.3\textwidth,angle=-90] {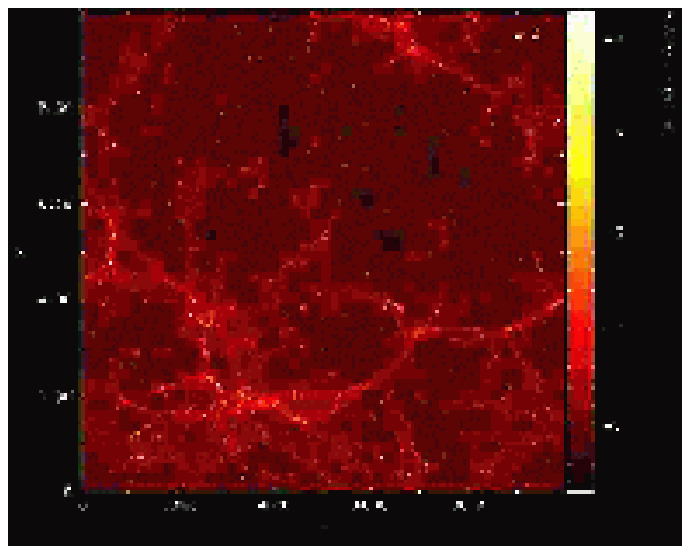}
}
\centerline{
  \includegraphics[width=0.3\textwidth,angle=-90] {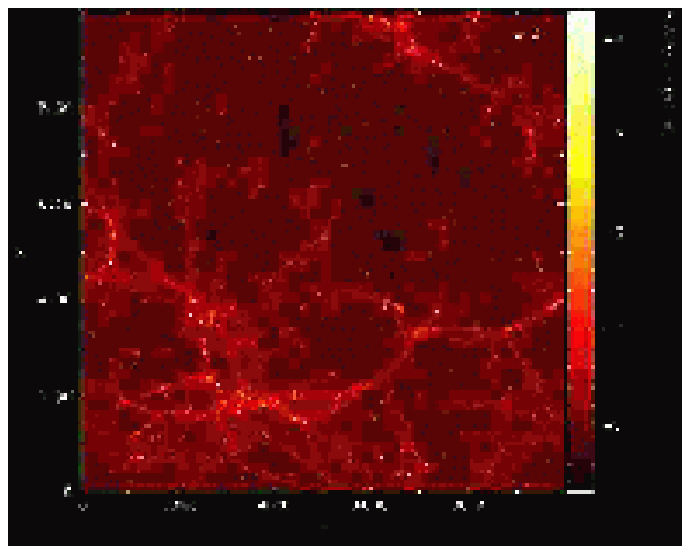}
  \includegraphics[width=0.3\textwidth,angle=-90] {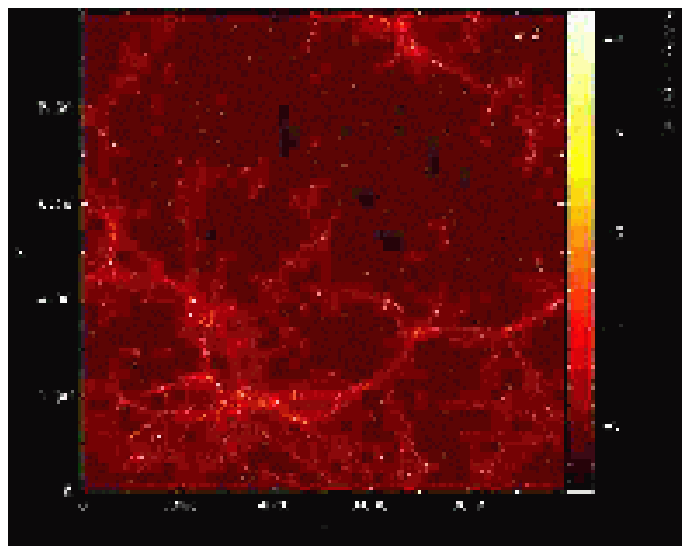}
  \includegraphics[width=0.3\textwidth,angle=-90] {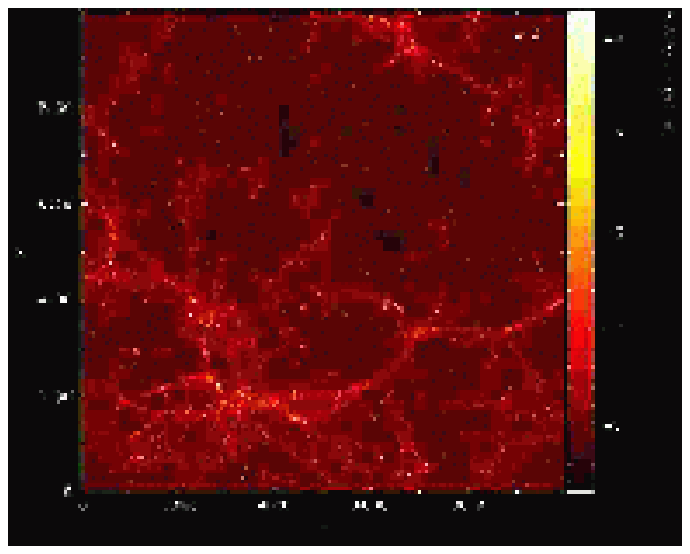}
}
\caption{
Gas column density at $z$=3 projected along with $z$-axis for the CW, NW, ME, 1.5ME, EE, and 1.5MM run from the top-left to bottom-right.
All panels use the same colour scale as shown in the side bar.
The size of each panel is comoving 10\,$h^{-1}$Mpc on the side. 
The gas distribution is almost identical in all the runs.
The visualisation was done by the SPLASH code with 3-D rendering mode \citep{Splash}.
}
\label{fig:snap_rho}
\end{figure*}

\begin{figure*}
\centerline{
  \includegraphics[width=0.3\textwidth,angle=-90] {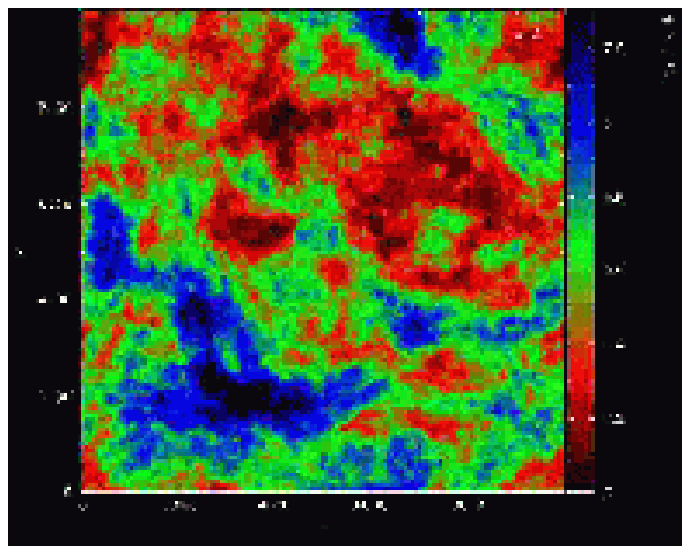}
  \includegraphics[width=0.3\textwidth,angle=-90] {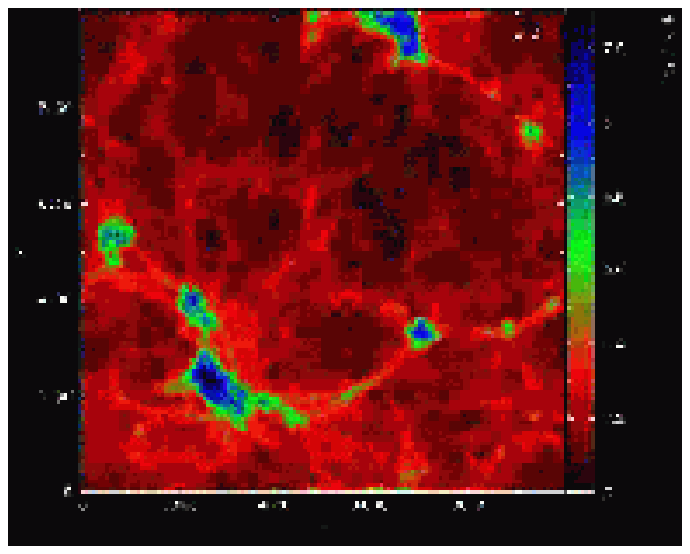}
  \includegraphics[width=0.3\textwidth,angle=-90] {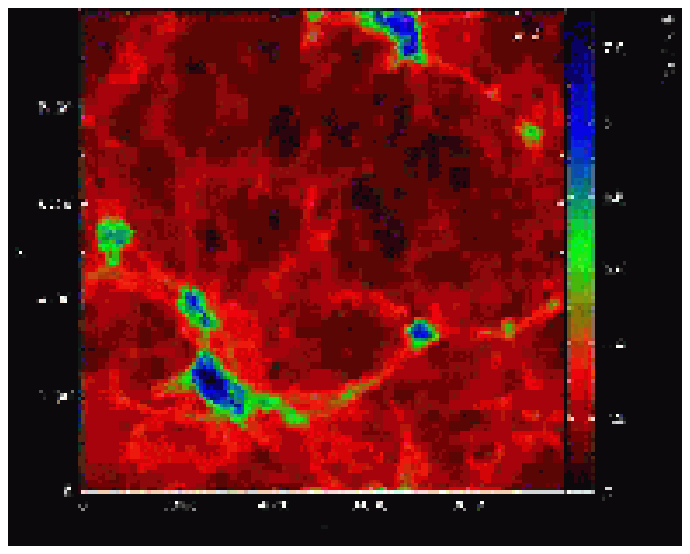}
}
\centerline{
  \includegraphics[width=0.3\textwidth,angle=-90] {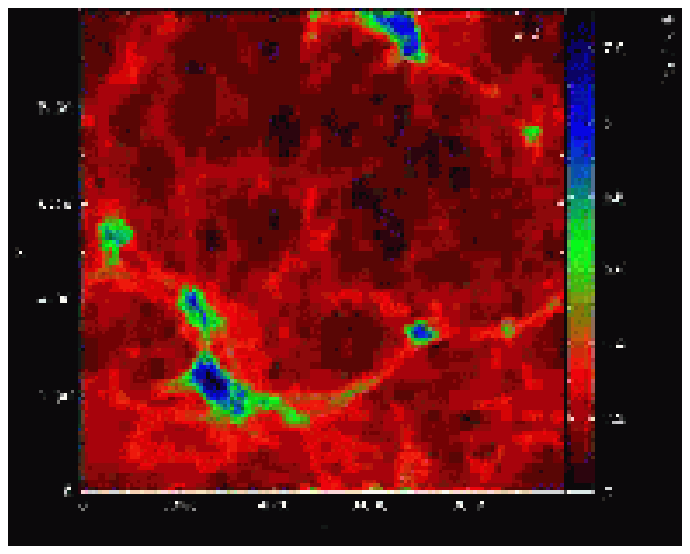}
  \includegraphics[width=0.3\textwidth,angle=-90] {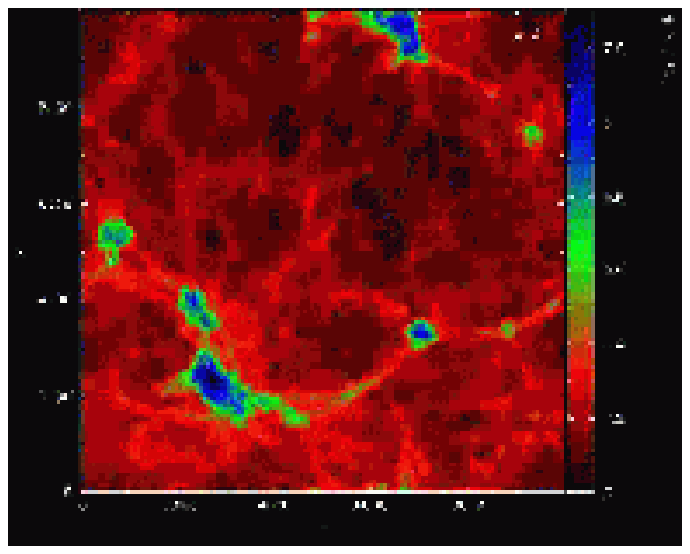}
  \includegraphics[width=0.3\textwidth,angle=-90] {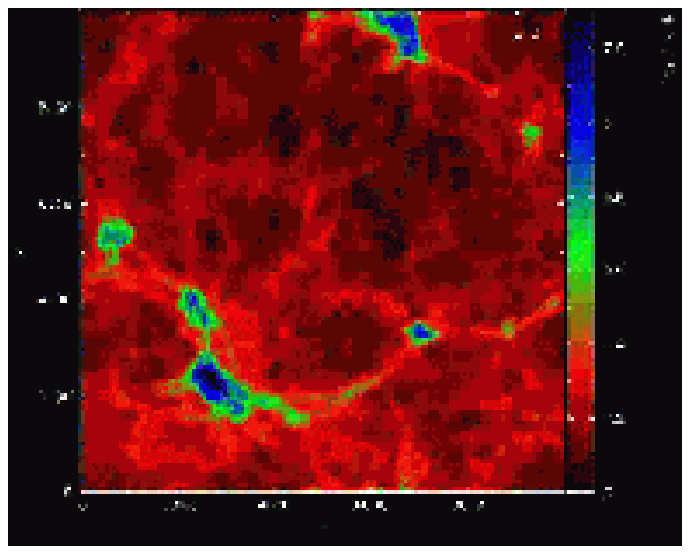}
}
\caption{
Same as in Figure~\ref{fig:snap_rho}, but for the projected internal energy ($\int u dz$): CW, NW, ME, 1.5ME, EE, and 1.5MM run from the top-left to bottom-right.
The IGM temperature is significantly higher on much larger scales in the CW run (top-left) than in the other runs. 
The MVV runs hardly heat the IGM on large scales. 
}
\label{fig:snap_u}
\end{figure*}

\begin{figure*}
\centerline{
  \includegraphics[width=0.3\textwidth,angle=-90] {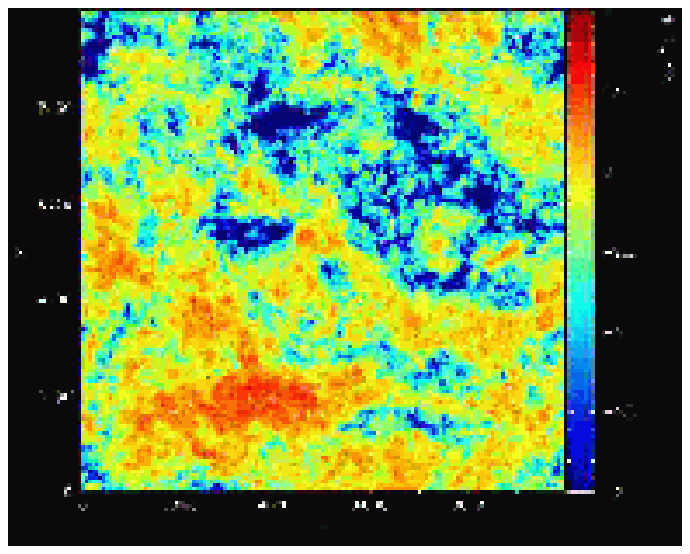}
  \includegraphics[width=0.3\textwidth,angle=-90] {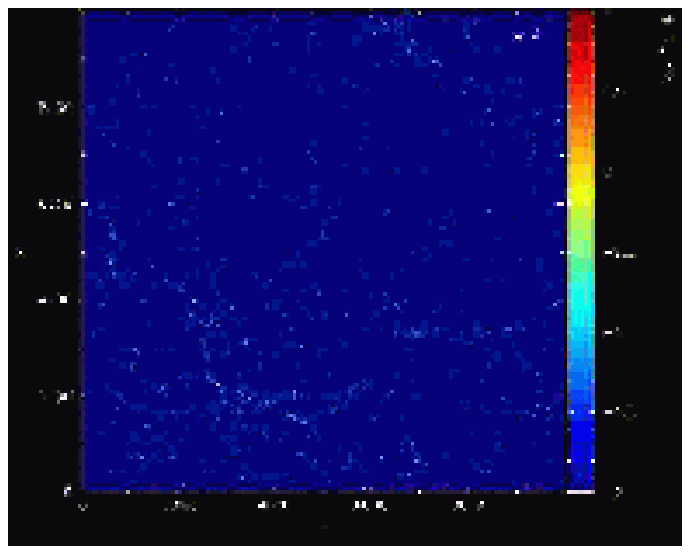}
  \includegraphics[width=0.3\textwidth,angle=-90] {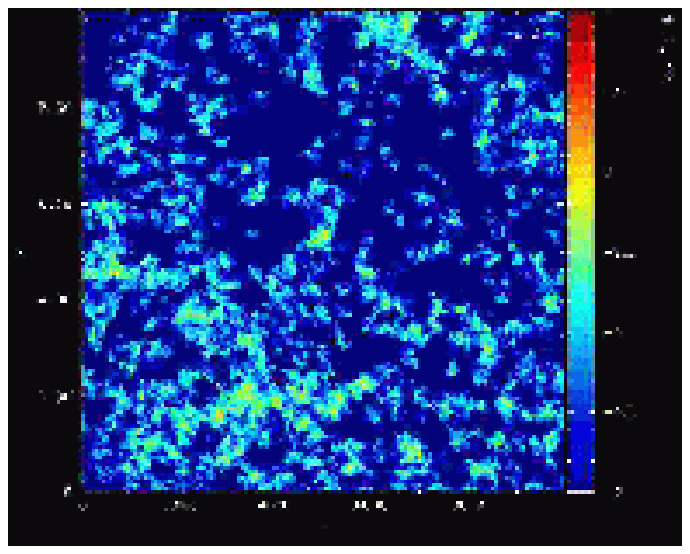}
}
\centerline{
  \includegraphics[width=0.3\textwidth,angle=-90] {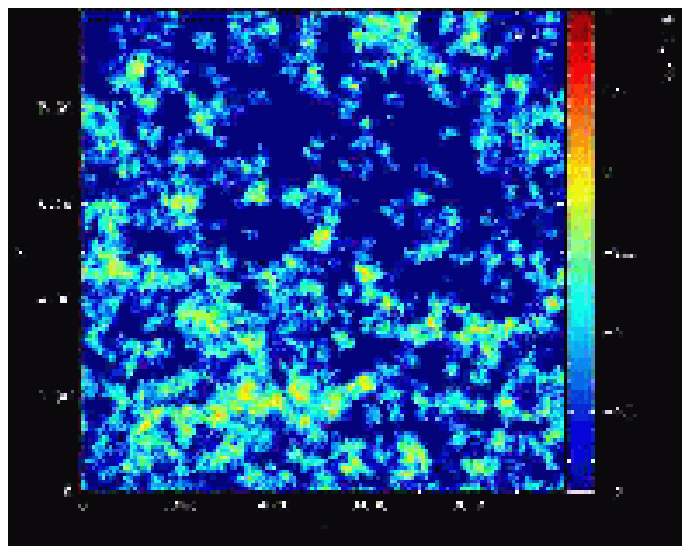}
  \includegraphics[width=0.3\textwidth,angle=-90] {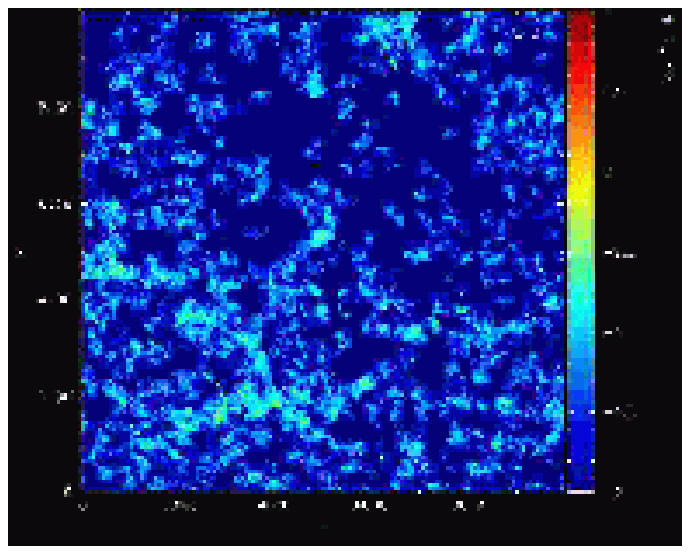}
  \includegraphics[width=0.3\textwidth,angle=-90] {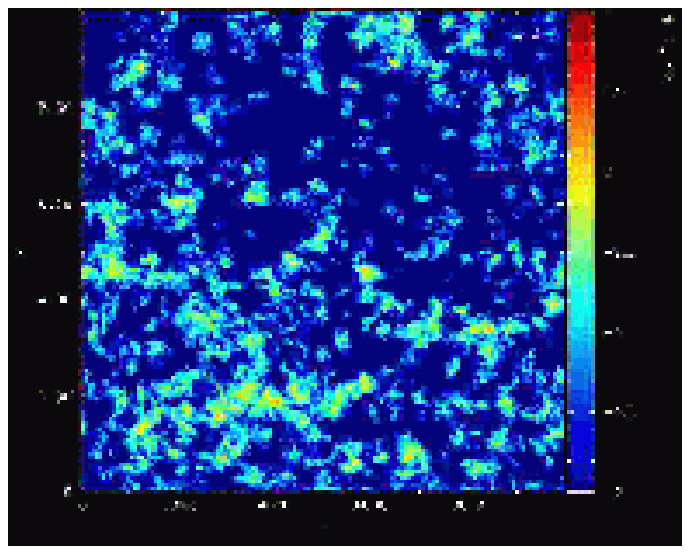}
}
\caption{
Same as in Figure~\ref{fig:snap_rho}, but for the projected metallicity ($\int Z dz$): CW, NW, ME, 1.5ME, EE, and 1.5MM run from the top-left to bottom-right.
The metals are spread over much broader regions in the CW run than in the other runs. 
The MVV runs also spread a significant amount of metals over the IGM, while the NW run (top middle panel) hardly spreads metals.
}
\label{fig:snap_Z}
\end{figure*}

Figures~\ref{fig:snap_rho} -- \ref{fig:snap_Z} show the snapshot view of the simulations with six different wind models at $z$=3: the CW, NW, ME, 1.5ME, EE, and 1.5MM run from top-left to bottom right. 
Figure~\ref{fig:snap_rho} shows that the gas column density distribution is very similar in all the runs, which implies that the amount of outflow material is negligible compared to the total gas content in the entire simulation box.

In contrast, the effect of outflow on the IGM temperature and the global metal distribution is very significant.
Figure~\ref{fig:snap_u} shows that there is a clear distinction between the CW run and the others in the distribution of gas thermal energy.
In the CW run, the IGM is significantly heated and the internal energy is much higher on larger scales than the other runs.
The similar energy distribution between the NW run and the MVV runs suggests that the MVV wind models can generate galactic outflows without overheating the IGM, not as much as the CW run.
This is an encouraging feature for the MVV wind model, because \citet{Oppenheimer.Dave:06} showed that the IGM temperature in the CW run is too high compared to the observed IGM temperature measured from the line-width of the Ly-$\alpha$ forest \citep{Schaye.etal:00}.
We will discuss this issue quantitatively in Section~\ref{sec:temperature}.

Figure~\ref{fig:snap_Z} shows the projected metallicity distribution for all the runs at $z=3$.
The CW run spreads the metals to the greatest distances, whereas the NW run spreads almost no metals over the intergalactic space.
This suggests that the mechanisms other than the galactic outflow, such as ram pressure stripping, are inefficient in enriching the IGM.
The MVV runs spread a significant amount of metals over the IGM, but not as much as the CW run. 
These snapshots demonstrate that the MVV wind can enrich the IGM without overheating it.

Among the four MVV wind models, the 1.5ME run shows the widest metal distribution.
The comparison between the 1.5ME and the ME run tells us that the high $\vwind$ ($\zeta =1.5$) causes a wider metal distribution.
The comparisons between the EE, ME, and 1.5MM run demonstrate the effect of mass-loading parameter $\eta$: the momentum-driven wind is overall more efficient in the IGM metal enrichment than the energy-driven wind.
The energy-driven wind tends to eject more winds from lower mass galaxies, which have lower metallicity, than the momentum-driven wind, owing to the $\siggal$ dependence of $\eta$.
The lower metallicity of the winds from lower mass galaxies gives rise to the less metal spread in the energy-driven wind than the momentum-driven wind.
In the MVV wind models, the choices of $\eta$ and $\vwind$ alter the degree of IGM metal enrichment.

\subsection{Cosmic Star Formation History}

Figure~\ref{fig:sfr} shows the cosmic star formation history of all the runs with different wind models.
The MVV wind runs show a significant suppression of SFR compared to the NW run, but a similar to or a higher SFR than the CW run.
Observations suggest that the range of cosmic SFR at $z \sim 3$ is between $0.07 - 0.3$\,$\Msun$\,yr$^{-1}$\,Mpc$^{-3}$ \citep{Hopkins.Beacom:06}, although there is a huge uncertainty \citep{Kistler.etal:09,Choi.Nagamine:09b}.
The comparison with the observed cosmic SFR shows that the NW run generates too many stars, while the results of the MVV wind runs are in the observed range.

The differences in cosmic SFR among the MVV runs yield insights on the effects of wind parameters ($\eta$ and $\zeta$).
Among the MVV runs, the EE run shows the lowest SFR and a good agreement with that of the CW run.
As discussed above, the energy-driven wind tends to produce galactic outflows more in lower mass galaxies.
Since low-mass galaxies form earlier in the hierarchical galaxy formation model, the energy-driven wind produces more outflows in the earlier epoch of galaxy formation.
Consequently, the energy-driven wind is more efficient in suppressing SFR than the momentum-driven wind.
The 1.5ME run suppresses the SFR more significantly than the ME run, because of the higher value of $\vwind$.

Interestingly, the cosmic SFR in the MVV runs does not show a turn-over at $z>3$. 
It implies that the peak of the cosmic SFR is located at $z < 3$ for the MVV runs, while the CW run has a clear peak at $z > 3.5$.
As shown in Figure~\ref{fig:snap_u}, the CW model overheats the IGM, which prohibits the gas accretion onto galaxies and suppresses star formation.
On the other hand, the MVV wind model does not heat the IGM as much as the CW model, and allows continuous IGM accretion that increases the cosmic SFR.
Therefore, the MVV wind model can shift the peak of the cosmic SFR to a lower redshift without changing the star formation model \citep[see][for the effect of star formation model on the cosmic SFR]{Choi.Nagamine:09b}.
In order to definitely locate the peak of the cosmic SFR, we need to run simulations with a larger volume which reach lower $z$.
Our current study suggests that the MVV wind model changes the shape and the amplitude of cosmic SFR.

\subsection{Galaxy Stellar and Baryonic Mass Functions}

Figure~\ref{fig:MF} shows the stellar and baryonic (i.e., star$+$gas) mass functions of the six different wind models at $z = 5$, 4, and 3.
In general, the stellar mass functions of MVV runs are located in-between the CW and NW run.
The trend is consistent with the cosmic SFR; the CW run generates the least amount of stars, while the NW run generates too many stars.
The consistency between the cosmic SFR and the galaxy mass function also holds among the MVV runs: the number of galaxies in the EE run is the lowest, and that of the ME run is the highest.
As discussed above, the choices of $\eta$ and $\vwind$ parameters determine the amount of star formation, as well as the stellar masses of galaxies.

Interestingly, the low-mass-end slope of both stellar and baryonic mass functions in the MVV runs are flatter than the other runs.
For example, the power-law slope is $\alpha \sim -1.6$ (assuming the Schechter function) in the mass range of $M_{\rm star} \sim 10^8 - 10^9 \Msun$ for the MVV run, while $\alpha \sim -2$ in the CW and NW run at $z = 3$. 
The number of low-mass galaxies is reduced in the MVV runs compared to the CW run, because the values of $\eta$ are higher for lower mass galaxies in the MVV runs, while the CW run adopts a constant value of $\eta$.
Therefore, the low-mass galaxies in the MVV runs lose comparatively more mass than those in the CW run (see also Section~\ref{sec:gfrac}).
Furthermore, less IGM heating in the MVV runs allows continuous gas accretion onto more massive galaxies, and the galaxies with $M_{\rm star} > 10^9 \Msun$ in the MVV run become more massive than those in the CW run.

\subsection{Gas Fraction of Galaxies}
\label{sec:gfrac}

Figure~\ref{fig:GFrac} shows the gas fraction of galaxies at $z = 5$ and 3.
In general, the gas fraction in the MVV run is roughly constant as a function of galaxy mass for galaxies with $M_\star = 10^8 - 10^9 \Msun$.  
At $z = 5$, the gas fractions of low-mass galaxies in the MVV runs are lower than those of the CW and NW run, and higher for massive galaxies.
Several mechanisms are at work to give rise to this result, but the dependency of $\eta$ on galaxy mass plays a key role, because it is directly related to the gas removal rate. 
Compared to the CW run, the MVV runs have higher $\eta$ and more gas is ejected from lower mass galaxies. 
Compared to the NW run, the MVV runs have lower SFRs, therefore they consume less gas in massive galaxies, resulting in higher $f_{\rm gas}$ in massive galaxies.
Among the MVV runs, the EE run shows the lowest gas fraction for the lower mass galaxies and highest gas fraction for the higher mass galaxies, because $\eta$ is the greatest in the EE run for low-mass galaxies and smallest for massive galaxies among different MVV runs. 
Note that not only $\eta$ but also $\vwind$ characterise galactic outflows.
However, $\vwind$ mostly affects the IGM properties.
Therefore, we argue that the difference in $f_{\rm gas}$ mainly comes from the $\eta$, which is more closely related to the gas removal rate. 

\begin{figure}
\centerline{\includegraphics[width=1\columnwidth,angle=0] {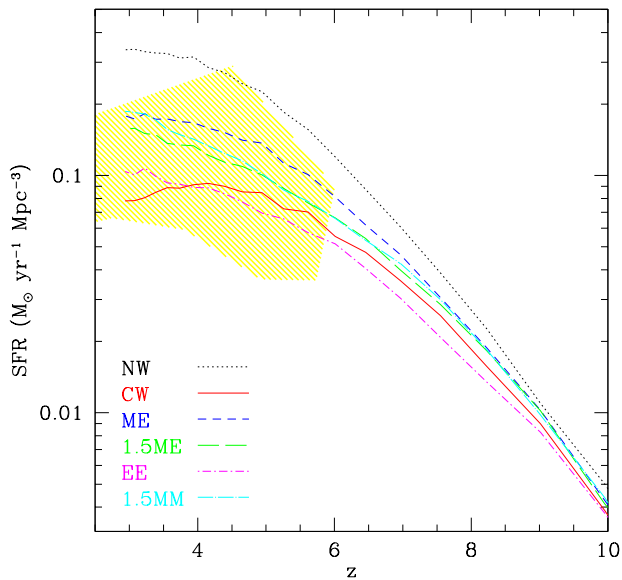}}
\caption{
Cosmic star formation history of the simulations with six different wind models.
The MVV wind runs have significantly lower SFRs compared to the NW run.
The CW and EE run have the lowest cosmic SFR.
For comparison, the yellow shaded region indicates the locus of the observed 
data points compiled by \citet{Nagamine.etal:06}. 
}
\label{fig:sfr}
\end{figure}

\begin{figure*}
\centerline{\includegraphics[width=1\textwidth,angle=0] {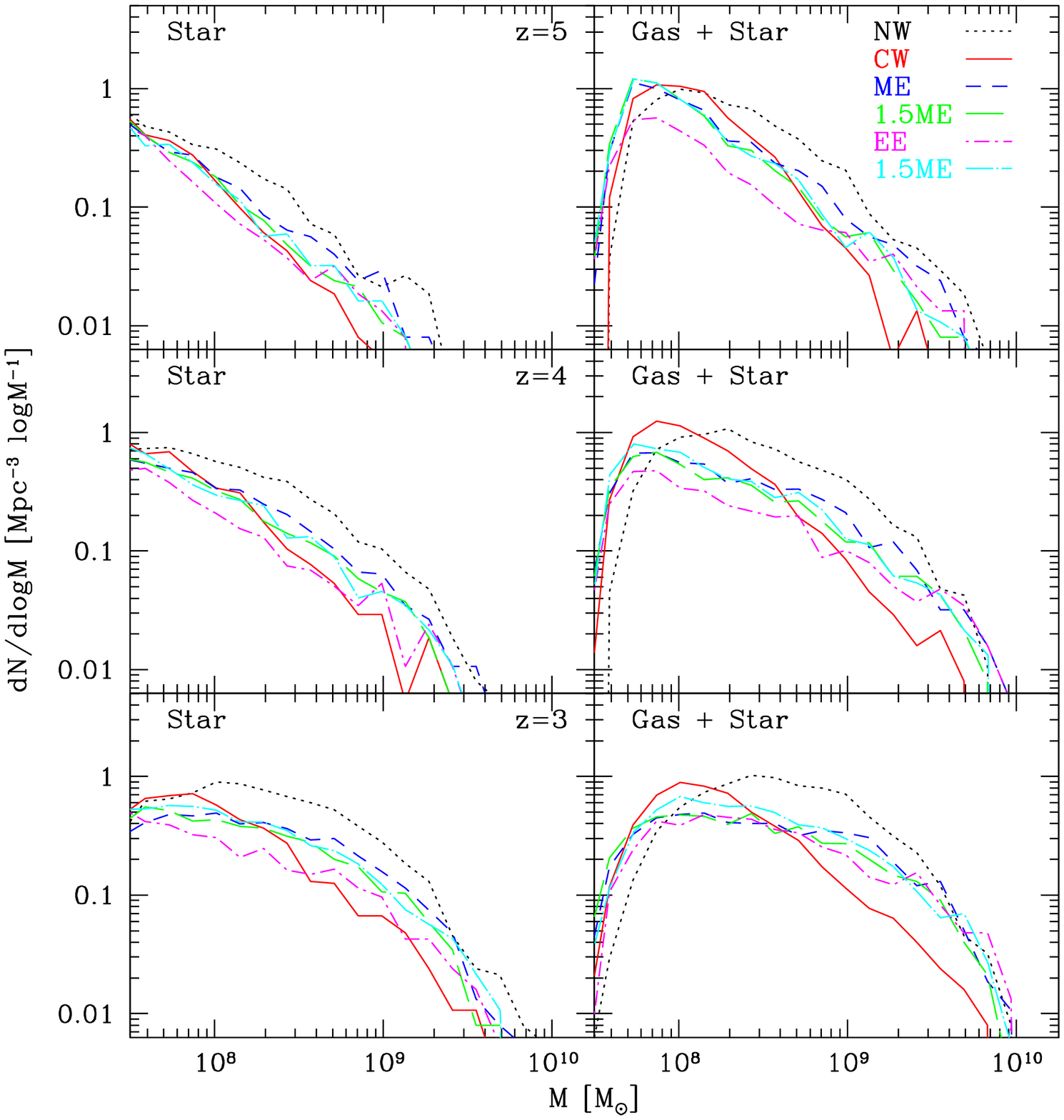}}
\caption{
Evolution of galaxy mass functions for the simulations with six different wind models.
The left panels are the stellar mass functions, and the right panels are the baryonic (star + gas) mass functions.
The mass functions in the MVV runs are flatter than that in the CW and NW runs.
}
\label{fig:MF}
\end{figure*}

\begin{figure*}
\centerline{\includegraphics[width=1\textwidth,angle=0] {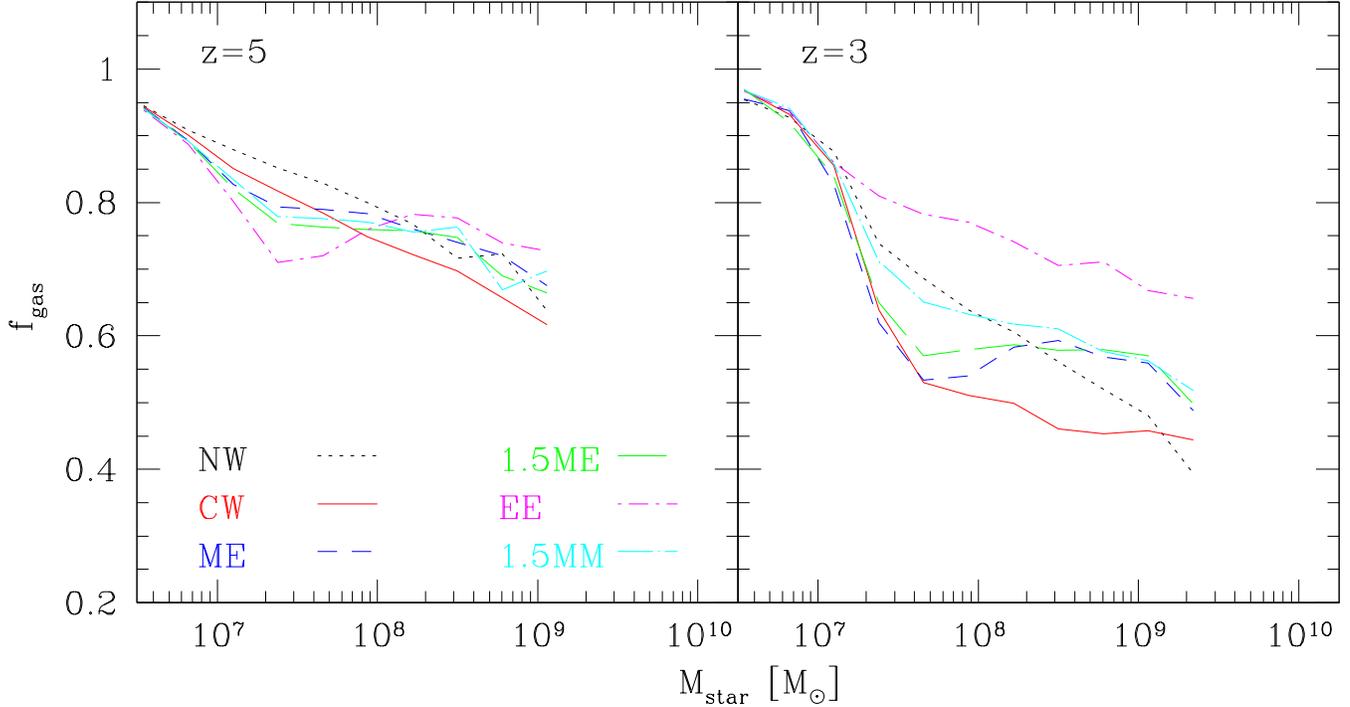}}
\caption{
Median gas fraction of galaxies as a function of galaxy stellar mass for six different wind models.
In general, $f_{\rm gas}$ in the MVV run is lower for the low-mass galaxies than in the CW and NW run, and roughly constant for galaxies with $M_\star = 10^8 - 10^9 \Msun$.
}
\label{fig:GFrac}
\end{figure*}

At $z = 3$, the general trend is similar to $z=5$. 
However, at lower redshifts, the balance between outflow and IGM accretion become more important for the gas fraction distribution.
At $z = 3$, the CW run has the lowest $f_{\rm gas}$, because its high IGM temperature reduces the IGM accretion onto galaxies.
One interesting feature in the $z = 3$ result is that the EE run shows the highest $f_{\rm gas}$.
This is because the SFR is the lowest in the EE run as we saw in Figure~\ref{fig:sfr}, and more gas is preserved in galaxies. 
The IGM temperature is also lower than the CW run, enhancing the IGM accretion onto galaxies. 
We will also see the enhanced IGM accretion in the EE run compared to the CW run in Section~\ref{sec:Phase}.

\subsection{Wind velocity}
\label{sec:vwind}

\begin{figure*}
\centerline{\includegraphics[width=1\textwidth,angle=0] {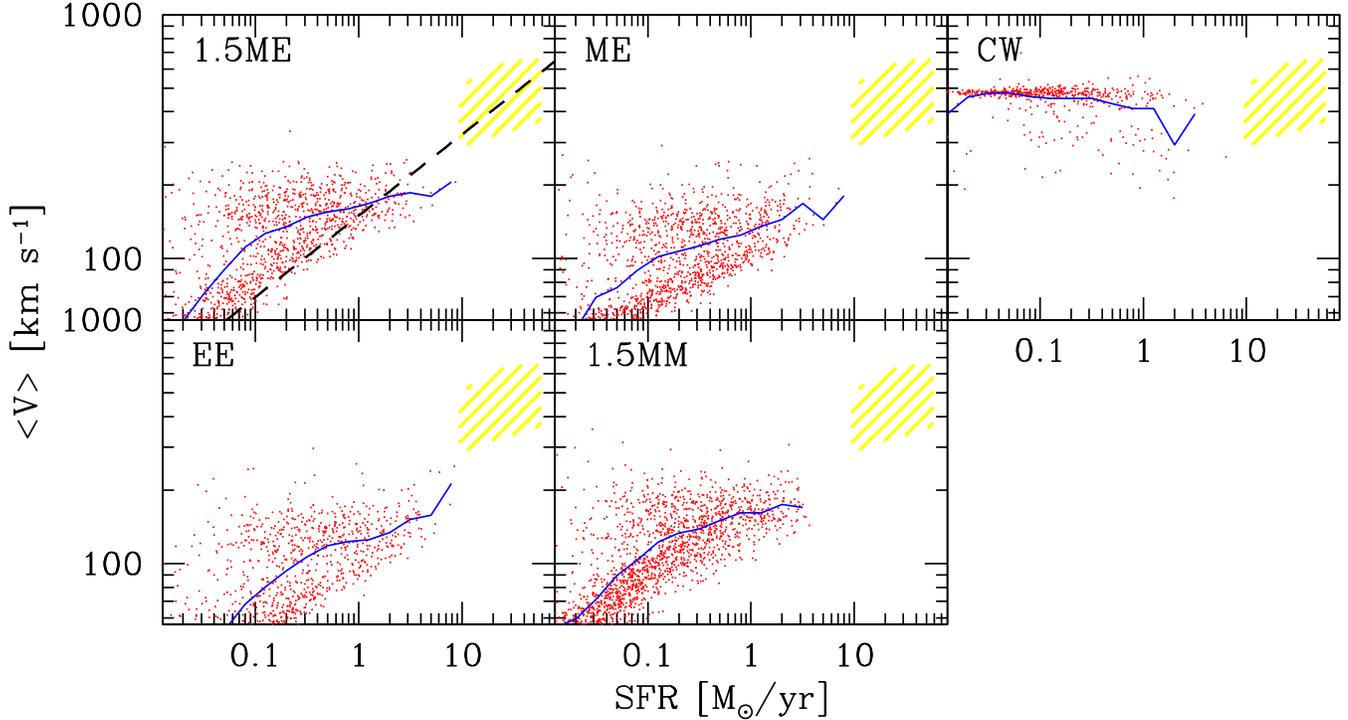}}
\caption{
The averaged outflow speed as a function of galaxy SFR.
For each galaxy, we measure the wind velocity of all wind particles and compute the average wind velocity, as represented by the red dots. 
The blue solid lines are the average speed for galaxies within a given mass range.
The black dashed line is the input wind velocity with $\zeta = 1$ at $z =$3, which is identical to the $z =$ 3 line on left panel of Figure~\ref{fig:windmodel}.
The yellow shade is the observed wind velocity range from \citet{Weiner.etal:09}.
}
\label{fig:sfr_v}
\end{figure*}

Figure~\ref{fig:sfr_v} shows the wind velocity as a function of galaxy SFR.
We directly measure the velocities of the wind particles for each galaxy, and then average them over for each galaxy.  
We assume that the wind particles were ejected from the nearest galaxy. 
This assumption may lead to some wrong associations between the wind particles and the launching galaxy, but the fraction of such mis-identification must be small and should not affect the average velocity. 
The figure shows that the wind velocity of MVV runs is a function of SFR, in good agreement with the input velocity (Eqns.~[\ref{eq:vwind}, \ref{eq:SFR_vesc}]; indicated by the long-dashed line in the top left panel), while the wind velocity of the CW run is independent of galaxy SFR.
Both the MVV and CW runs show a non-negligible scatter, which results from the velocity dispersion of gas particles in a galaxy prior to becoming a wind.

We compare our results with the observed wind velocity by \citet{Weiner.etal:09}. 
Owing to the flux limit, the observation only probes the wind velocity of massive galaxies with SFR greater than 10\,$\Msun$\,yr$^{-1}$. 
The observations of wind velocities also contain significant scatter, and it is difficult to distinguish a favored model \citep{Martin:05,Weiner.etal:09}. 
However, many observations have suggested that the wind velocity is a function of galaxy size with a similar relationship to our model, and the results of the MVV run is consistent with the limited observations of wind velocities. 

The constant wind velocity in the CW run causes the high IGM temperature (see Figure~\ref{fig:snap_u} and Section~\ref{sec:temperature}).
However, this difference in the IGM temperature does not result from different outflow energy.
The injected kinetic energies depend on both the outflow velocity and the amount of the outflow mass, which depends on $\eta$. 
We find that the injected kinetic energies by winds in the CW run and the MVV runs for a given galaxy SFR are indifferent, because low-mass galaxies in the CW run generate high-velocity outflow, while the low-mass galaxies in the MVV runs have higher mass loading factors. 
What makes a big difference in the IGM temperature is the high wind velocity from low-mass galaxies. 
Since the potential wells of these galaxies are shallow, their winds propagate to fill a significant volume, and their effect on the environment is significant. 
In contrast, the winds from low-mass galaxies in the MVV runs do not have such a high velocity, so their effect on the environment is limited.
The high IGM temperature in the CW run conflicts with the observed IGM temperature. 

\section{Evolution of IGM}
\label{sec:IGM}

\subsection{Phase Evolution}
\label{sec:Phase}

\begin{figure*}
\centerline{\includegraphics[width=1\textwidth,angle=0] {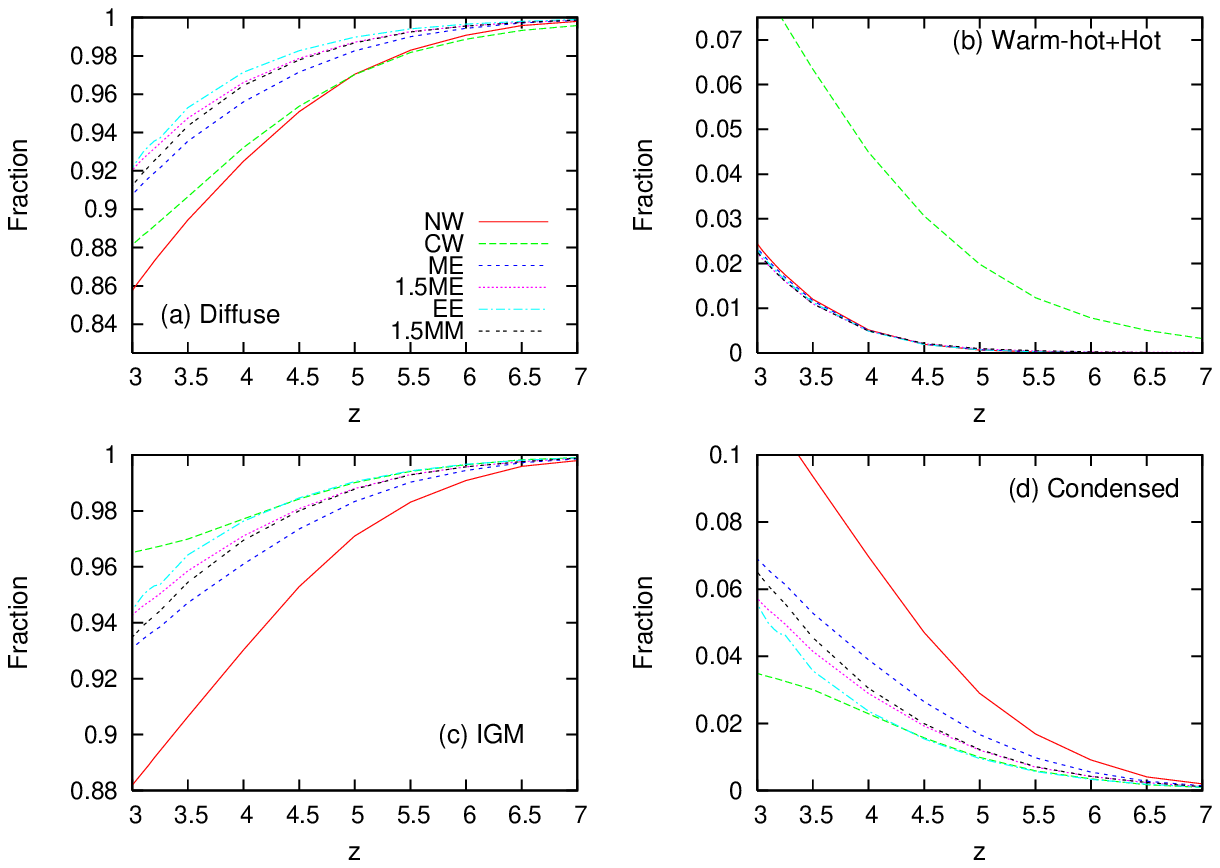}}
\caption{
Evolution of the mass fractions of the four different baryon phases (hot, warm-hot, diffuse, and condensed) for the simulations with six different wind models.
{\it Panel (b)} shows the addition of the two phases, `warm-hot + hot', and {\it panel (c)} shows the total mass in the IGM, i.e., `diffuse + warm-hot + hot'.
This figure shows that the outflow medium in the MVV wind model mostly goes into the `diffuse' phase instead of the `warm-hot' phase, whereas in the CW run, it is the opposite.
}
\label{fig:Evolv4}
\end{figure*}

Figure~\ref{fig:Evolv4} shows the evolution of mass fractions of different baryon phases.
The baryons in the Universe can be broadly categorised into four different phases according to their overdensity and temperature: `hot', `warm-hot', `diffuse', and `condensed' \citep{Dave.etal:99,Cen.Ostriker:99,Dave.etal:01}.  
The 'hot' phase is the non-starforming gas with $T > 10^7$\, K.
The 'warm-hot' phase is the non-starforming gas with $10^5 < T < 10^7$\, K.
The 'diffuse' phase is non-starforming gas with $T < 10^5$\,K and $\delta < 1000$, where $\delta = \rho/\bar{\rho}-1$ and $\bar{\rho}$ is the mean baryonic density.
The 'condensed' phase includes non-starforming gas with $T < 10^5$\,K and $\delta > 1000$ , as well as the star-forming gas and stars.
The 'hot' and 'warm-hot' phases are mostly the shock-heated gas in clusters and groups of galaxies.
The `diffuse' phase is mostly the photoionised IGM with lower temperature, which can be observed as the Ly-$\alpha$ forest.  
The `condensed' phase is the high density, colder gas in galaxies.

The {\it panels (a) and (b)} show that the MVV runs always contain more diffuse gas and less `warm-hot + hot' gas than the CW run.
This difference results from the velocity and the temperature of the outflowing gas.
After the wind particle is ejected, it is mixed with the ambient gas and the temperature can be decreased via adiabatic cooling.
Meanwhile, the ambient gas temperature increases due to the thermal energy from the outflowing gas and shock heating. 
As discussed in Section~\ref{sec:vwind}, the galactic outflow from low-mass galaxies in the CW run influences the ambient gas significantly. 
In general, the heating from these outflows increase `warm-hot + hot' gas fraction in the CW run.
The comparison of the CW and MVV run in {\it panels (a) and (b)} shows that most of the outflow medium goes into the `warm-hot + hot' phase in the CW run, whereas in the MVV runs, it goes into the `diffuse' phase.  
In conclusion, the MVV wind model significantly alters the mass fraction of different baryon phases.

The evolution of the `condensed' and `IGM' phase ({\it panels (c) and (d)}) is related to the cosmic SFR, because the amount of 'condensed' phase and SFR is closely correlated.
The CW run shows the highest fraction of the IGM and the lowest fraction of the condensed gas, while the NW run shows the opposite.
Among the MVV runs, the EE run has the highest fraction of the IGM, owing to the stronger outflows from low-mass galaxies.
We also see that the IGM mass fraction rapidly declines in the EE run at $z \lesssim 4$ compared to the CW run. 
This is because the low SFR produces weak galactic outflow and the low IGM temperature increases the IGM accretion onto galaxies in the EE run.
This increase of the IGM accretion is also responsible for the high $f_{\rm gas}$ for the galaxies in the EE run at $z = 3$ (see Section~\ref{sec:gfrac}.

\subsection{Temperature Evolution}
\label{sec:temperature}

\begin{figure}
\centerline{\includegraphics[width=1\columnwidth,angle=0] {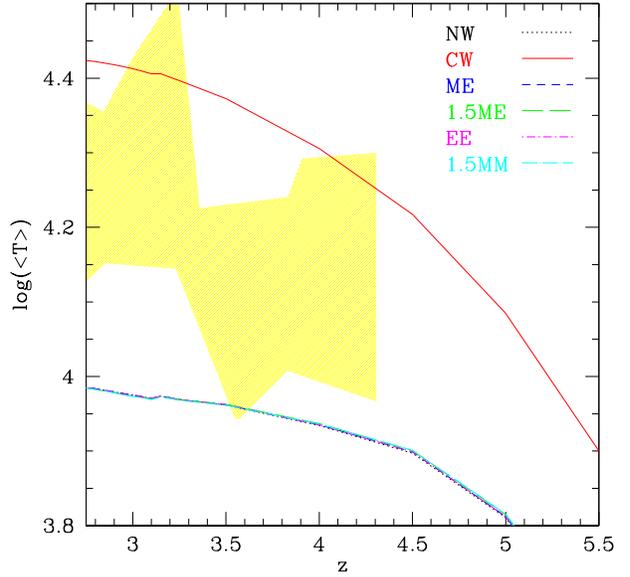}}
\caption{
Evolution of the volume-weighted temperature for six simulations with different wind models.  
Except the CW run, all simulations show almost identical temperature evolution.
The yellow shade is the range of observed IGM temperature derived by \citet{Schaye.etal:00} from the Ly-$\alpha$ forest line-widths.
}
\label{fig:IGM_T}
\end{figure}

Earlier, we showed in Figure~\ref{fig:snap_u} that the NW and MVV run hardly heat the IGM, while the IGM in the CW run is significantly hotter.
In order to quantitatively discuss this issue, we plot the evolution of volume-weighted temperature of all gas in the simulation box in Figure~\ref{fig:IGM_T}. 
Note that the overall IGM property is better represented by the volume-weighted value, because the IGM covers a large volume of the Universe including low-density voids and high-density filaments.
Figure~\ref{fig:IGM_T} confirms that the IGM temperature in the CW run is higher than the other runs at all redshifts.  
It confirms that the average IGM temperature in the MVV run is very similar to that of the NW run.  
Since the IGM heating by the outflows in the NW run should be minimal, this figure also shows that the MVV wind model indeed does not heat the IGM.

\citet{Schaye.etal:00} estimated the IGM temperature at the mean density using the Ly-$\alpha$ absorption line widths, as shown by the yellow shade in Figure~\ref{fig:IGM_T}.
At $ 3.5 \lesssim z \lesssim 4.5$, the results of the MVV runs are at the lower edge of the observed estimates. 
At $ z \lesssim 3.5$, the observed temperature range of $\approx 1.5 - 2.5 \times 10^4$\,K is higher than the MVV run results ($\sim 1 \times 10^4$\,K), and similar to or lower than the CW run result ($\sim 3 \times 10^4$\,K).
\citet{Schaye.etal:00} argued that the IGM temperature shows a peak and the gas becomes nearly isothermal at $z \sim 3$, which can be interpreted as an evidence for the He\,{\sc ii} reionization \citep[e.g.,][]{Sokasian.etal:03,McQuinn.etal:09}. 
The optically thin approximation is adopted for the UV background radiation in many cosmological simulations, however, this method underestimates the He\,{\sc ii} photoheating rate.
The mean distance of He Lyman limit systems is smaller than the mean distance between the sources of ionising photons, in which case the optically thin approximation is not appropriate, and a significant fraction of emitted ionising photon is absorbed and enhances the photoheating rate.
We need to consider the full radiative transfer effect in order to properly estimate the thermal history of IGM around He\, {\sc ii} reionization epoch \citep{Abel.Haehnelt:99}.
The neglect of this radiative transfer effects results in the IGM temperatures which are too low by a factor of 2 after He\,{\sc ii} reionization \citep{Jena.etal:05,Faucher-Giguere.etal:09}.
Therefore the IGM temperature  $ z \sim 3$ in simulations with an optically thin approximation should not be higher than the observations, if the assumed amplitude of UV background is reasonable. 
Our results suggest that the CW run clearly overheats the IGM, while the MVV runs do not.

We also note that the IGM temperatures in the CW and NW run are lower than that reported by \citet[][]{Oppenheimer.Dave:06}, despite their outflow prescription is very similar to ours.
This may result from the different cosmology such as lower $\sigma_{8}$ and lower $n_{s}$ that we adopt, or different metal cooling method from theirs. 
In particular, recently \citet{Shen.etal:09} argued that the metal cooling rate in \citet{Choi.Nagamine:09a} was overestimated compared to their more refined treatment, owing to the missing treatment of metal ionisation by the UV background effect.
We will investigate this issue in the future.

\subsection{Metal Enrichment}
\label{sec:enrichment}

\subsection{Overdensity -- Metallicity Relationship}

Metals are generated in galaxies that are usually located in dense regions, and spread over the low-density IGM. 
Therefore it is expected that there will be a positive correlation between gas density and metallicity.
Using the \Civ\ pixel optical depth statistics, \citet{Schaye.etal:03} estimated the relationship: ${\rm [C/H]} = -3,47 + 0.08(z-3)+0.65[\log(\delta)-0.5]$, where $z$ is the redshift range between 1.8 to 4.1, and $\delta \equiv \rho/\bar{\rho}-1$ is the baryonic overdensity.
This relationship indicates a positive correlation between the IGM metallicity and baryonic overdensity. 

Figure~\ref{fig:den_metal} shows the baryonic overdensity$-$metallicity relation at $z = 3$ for the simulations with six different wind models.
We also overplot the relationship derived by \citet{Schaye.etal:03} with its 1-$\sigma$ scatter (shown by the yellow shade).
The runs with galactic outflows overpredict the gas metallicity compared to the observational estimate, while the NW run underestimates the metallicity.
This suggests that the galactic outflow is necessary to reproduce the observed IGM metallicity, but also that the current outflow models are spreading the metals too efficiently.

In Figure~\ref{fig:den_metal}, the CW and NW run are definitely outside the 1-$\sigma$ range of the observed data.
The MVV runs generally show somewhat higher metallicity than the observed data, but in better agreement with the data than the CW and NW run. 
It implies that the MVV wind reduces the degree of metal spreading, more consistently with the observation than the CW or NW run. 
Among different MVV runs, the EE run shows better agreement with the observation, because in this run, the low-mass, low-metallicity galaxies have high mass-loading factors for the energy-driven wind, and the outflow medium tends to have a lower metallicity.
In addition, the cosmic SFR in the EE run is the lowest among different MVV runs.
The low SFR results in less metal production, which also contributes to the better agreement with the observed density--metallicity relationship for the EE run.

\subsection{\Civ\ Statistics}

\begin{figure}
\centerline{\includegraphics[width=1\columnwidth,angle=0] {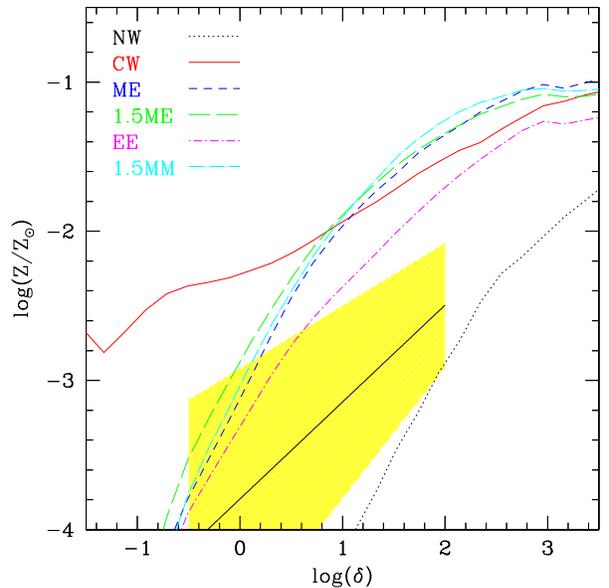}}
\caption{
Baryonic overdensity--metallicity relation at $z$=3 for six simulations with different wind models.  
The black solid line is the observational estimate derived by \citet{Schaye.etal:03} from the Ly-$\alpha$ forest line-widths, and the yellow shade is the 1-$\sigma$ scatter.
}
\label{fig:den_metal}
\end{figure}

The most commonly encountered metal lines associated with the Ly-$\alpha$ forest are the \Civ\ $\lambda \lambda$ 1548.2041, 1550.7812 doublet \citep{Cowie.etal:95,Ellison.etal:00}.
Therefore, many IGM metallicity studies use the \Civ\ measurements to estimate the total metallicity including \citet{Schaye.etal:03}.
\citet{Oppenheimer.Dave:06} argued that \Civ\ is not a straightforward measure of the IGM metallicity owing to the evolution of ionisation fraction, which is affected by the IGM temperature.
They also showed that the observed flat evolution of $\Omega$(\Civ) is caused by the evolution of ionisation fraction and the UV background.
Although $\Omega$(C) continuously increases, the increasing ionisation fraction compensates, and give rise to a flat evolution of $\Omega$(\Civ) evolution.
Therefore, comparing the observed IGM metallicity estimated from the \Civ\ measurements with the total metallicity in the simulation may not be a fair comparison, and this could be one of the reasons for the discrepancy between the observation and simulations shown in Figure~\ref{fig:den_metal}.
To avoid this complication, we directly compare the evolution of $\Omega$(\Civ) in our simulations with observations.

Owing to the computational cost, our current simulations do not track the evolution of individual metal element.
However we track the evolution of the total metallicity, $Z$, using the primordial metal ratio for $Z < 0.1 \Zsun$  \citep{Wheeler.etal:89,Bessell.etal:91}, and the solar abundance for $Z > \Zsun$ \citep{Anders.Grevesse:89}.
For $0.1 < Z/\Zsun < 1.0$, the abundance pattern is computed by interpolating between the primordial and the solar abundance patterns \citep[see][for detailed discussion]{Choi.Nagamine:09a}.
We estimated $\Omega$(C) from the total metallicity evolution with a given abundance pattern.
Abundance of \Civ\ is estimated from the abundance of carbon and its ionisation fraction.
We use the c96 version of Cloudy \citep{Cloudy} assuming an optically thin slab of gas with the ionisation radiation field of \citet{Haardt.Madau:96}.
The amplitude of the ionisation background is reduced by a factor of 1.6 to match the observed Ly-$\alpha$ flux decrement \citep{Oppenheimer.Dave:06}.

\begin{figure}
\centerline{\includegraphics[width=1\columnwidth,angle=0] {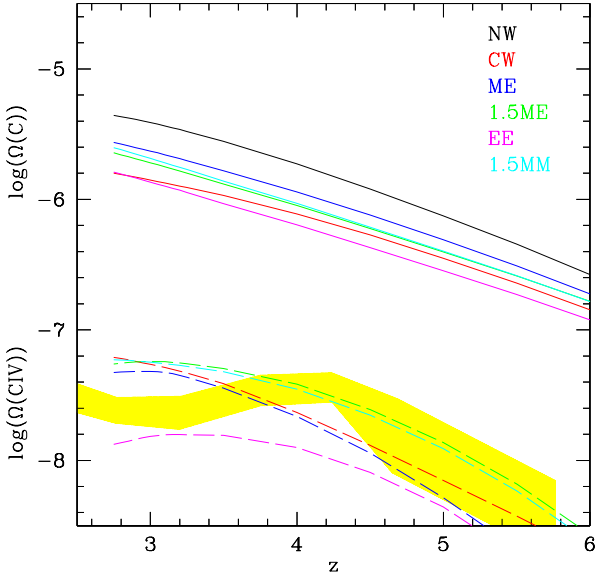}}
\caption{
Evolution of $\Omega$(C) (solid lines) and $\Omega$(\Civ) (dashed lines) as a function of redshift for all simulations with six different wind models. 
The yellow shade indicates the range of observed data of $\Omega$(\Civ) \citep{Songaila:01,Pettini.etal:03,Ryan_weber.etal:09}.
Our result of $\Omega$(\Civ) in the NW run is so low that it is located outside of the axis range.
}
\label{fig:c4}
\end{figure}

In Figure~\ref{fig:c4}, we show the redshift evolution of $\Omega$(C) and $\Omega$(\Civ), together with the observed range in yellow shade. 
In our simulations, $\Omega$(C) monotonically increases with decreasing redshift, as the total metallicity increases. 
The $\Omega$(\Civ) also continuously increases, except that the EE run turns over at $z\lesssim 3.3$. 
The MVV runs show a more mild evolution, and $\Omega$(\Civ) becomes flat at $z < 3.5$.
Observations show that the redshift evolution of $\Omega$(\Civ) is almost flat between $z = 2$ and 4.5, and decreases at $z > 5$ \citep{Ryan_weber.etal:09}.
Although the agreement is not perfect, the $\Omega$(\Civ) evolution in the simulations with galactic outflows is more or less consistent with the observed range (yellow shade). 
The NW run completely underestimates the $\Omega$(\Civ) by about an order of magnitude, therefore its result is not shown in the lower part of Figure~\ref{fig:c4}. 
This suggests that a galactic outflow model of some form is required to reproduce the observed $\Omega$(\Civ).
The flattening of $\Omega$(\Civ) at $z<3.5$ in the MVV runs is encouraging, and these runs may show a better agreement with future observations, if we extend the simulations to a lower $z$.

The discrepancies between our simulations and observations in Figure~\ref{fig:c4} is significantly less than those in Figure~\ref{fig:den_metal}. 
It suggests that the IGM temperature distribution might be responsible for the discrepancies in the overdensity--metallicity relation as we described earlier. 
These comparisons suggest that we need to test and improve the UV background model as well as the outflow model to resolve the discrepancies shown in Figures~\ref{fig:den_metal} and Figure~\ref{fig:c4}.

The evolution of $\Omega$(\Civ) has been used by several previous studies to verify galactic outflow models in cosmological simulations \citep[e.g.,][]{Oppenheimer.Dave:06}.
Compared to the Figure~12 of \citet{Oppenheimer.Dave:06}, the evolution of $\Omega$(\Civ) in our MVV runs show a similar level of agreement with observation.
Therefore our MVV wind model is also a legitimate new approach for modeling galactic outflow in cosmological simulations.

\section{Summary and Discussion}
\label{sec:summary}

In this paper, we developed a phenomenological multicomponent variable velocity (MVV) galactic outflow model for cosmological SPH simulations.
In this new wind model, the wind parameters such as $\eta$ and $\vwind$, are based on the galaxy mass and SFR, which are computed from the on-the-fly group-finder.

In the MVV wind model, all gas particles in a given galaxy have the same probability to become winds, and we allow lower density, cold gas to be part of the wind, which was not allowed in the previous models.
This allows our MVV wind to reflect the fact that wind material can arise from different phases in the ISM, and the MVV wind significantly enriches the IGM metallicity without overheating it, unlike the previous CW model. 
The comparison of our simulations with the observed overdensity--metallicity relationship shows that the IGM metallicity in our simulation is slightly higher than the observations.
However, our simulations show a reasonable agreement with the observational estimates of $\Omega$(\Civ), though not perfect. 
Because current observations rely only on a few metal absorption lines such as \Civ, the effects of UV background radiation and the IGM temperature may be responsible for this discrepancy, and we need to improve the treatment of UV background in the future. 

Recent WMAP measurement of the electron scattering optical depth \citep{Dunkley.etal:09} and the Ly-$\alpha$ forest transmission \citep{Chiu.etal:03} suggest that the UV background radiation model of \citet{Haardt.Madau:96} may need to be improved.
In particular, \citet{Faucher-Giguere.etal:09} investigated the implications of Ly-$\alpha$ forest opacity measurements at $2 \lesssim z \lesssim 4$, and found a remarkably flat ionisation rate over this redshift range.
\citet{Hambrick.etal:09} examined the effects of different ionisation rate evolution as a function of redshift on galaxy formation and evolution.  
They found that increase in either the intensity or hardness of ionising radiation, which recent input UV background radiation models predict comparing to old models, generally reduces early star formation and push it toward lower redshifts.
This change also results in enhanced late gas inflows, decrease of stellar half-mass radii, increase of central velocity dispersion, and reduction of substructures.

The MVV winds effectively remove the material from galaxies into the IGM.
In the runs with the MVV wind model, the cosmic SFR is significantly reduced compared to the NW run, and the results are more consistent with the observed cosmic SFR.
Because of the variable wind velocity and the functional dependence of $\eta$ on $\siggal$, the galaxy stellar mass functions in the MVV runs are flatter than that in the CW and NW runs.
Owing to the small volume of the simulations in this paper, we are only able to demonstrate this effect for the high-$z$, low-mass galaxies, and the flux limit of current observations do not reach this faintness yet.
Nevertheless, the flat galaxy stellar mass function at the low-mass end may provide a better agreement with observations for the low-$z$, intermediate mass galaxies, which can be tested by our future simulations with a larger volume.

Although our new MVV wind model has new features and show a good agreement with observations, there are still many uncertainties in the details of the wind physics.
For example, it is not clear when, where, and how the momentum-driven wind and the energy-driven wind dominates over one another.
In the MVV wind models, we mixed the momentum-driven wind and the energy-driven wind based on the local gas density. 
This mixture method and criterion are based on a logical but an ad hoc argument, and they need to be carefully evaluated and extensively tested.
From the comparisons with observations presented in this paper, it is still difficult to distinguish different models and identify a particularly favoured model.
The overdensity--metallicity relation (Figure~\ref{fig:den_metal}) suggests that the EE run is the favoured model, but the $\Omega$(\Civ) evolution shows that the EE run presents the worst agreement among different MVV runs.
Despite of these uncertainties, the comparisons among the four MVV models provide some insight into the wind models.
The result from EE run shows that it can reduce the SFR considerably, but this leads to significantly increased gas accretion at $z \lesssim 4$.
It suggests that the energy-driven wind is a possibly legitimate mechanism for low mass and high-$z$ galaxies, but a stronger outflow model is required for high-mass, low-$z$ galaxies.

Also, the details of driving mechanism of the multiphase outflow and its thermal composition are not well understood.
For example, the observed values of $\vwind$ and its functional dependence on various galaxy parameters, such as the stellar mass or SFR, are still unclear. 
We postulated $\vwind \propto (SFR)^{1/3}$ based on Equations (\ref{eq:vc}) and (\ref{eq:SFR_M}).
However, it is well known that $v_c \propto \Mstar^{1/4}$ for early type galaxies \citep{Faber.Jackson:76}, where $\Mstar$ is the galaxy stellar mass.
\citet{Noeske.etal:07} showed a strong positive correlation of $SFR \propto \Mstar$.
Combining the above two relationships suggests $\vwind \propto (SFR)^{1/4}$.
In this paper, we adopt the relationship $\vwind \propto (SFR)^{1/3}$ based on the arguments presented in Section~\ref{sec:MVV}, and also because some observations prefer $\vwind \propto (SFR)^{1/3}$ scaling \citep{Martin:05,Weiner.etal:09}.

Given these current uncertainties in both theory and observations, it is still somewhat premature to constrain and determine the best model among the four tested MVV wind models.  
Therefore, our new model may be yet another approach to the phenomenological model of galactic wind in cosmological SPH simulations, but our goal is to develop a better model for star formation and feedback incrementally for galaxy formation simulations. 
To further discriminate the models, we intend to perform further analyses of our simulations with regard to the quasar absorption lines, e.g., Ly-$\alpha$ forest and damped Ly-$\alpha$ absorbers.
At least the current comparisons between our simulations and observations provide some improved understanding on the role of $\eta$ and $\vwind$ in determining the cosmic SFR and IGM metal enrichment.

In summary, our new galactic outflow model, which reflects the fact that the wind material can arise from different phases of ISM and captures the dependence of variable wind velocity on galaxy mass and SFR, provides better agreement with various observations, such as the cosmic SFR, gas overdensity--metallicity relationship, and $\Omega$(\Civ).
These comparisons suggest that the MVV wind model is definitely favoured over the CW model, and that it is a more viable choice as a wind model for the future simulations. 


\section*{Acknowledgements}

We are grateful to Volker Springel for providing us with the original version of GADGET-3, on which our simulations are based, as well as those in \citet{Choi.Nagamine:09a, Choi.Nagamine:09b}.
We also thank Benjamin Oppenheimer for his help on our \Civ\ calculation.
This work is supported in part by the NSF grant AST-0807491, National Aeronautics and Space Administration under Grant/Cooperative Agreement No. NNX08AE57A issued by the Nevada NASA EPSCoR program, and the President's Infrastructure Award from UNLV. 
This research is also supported by the NSF through the TeraGrid resources provided by the Texas Advanced Computing Center.
Some numerical simulations and analyses have also been performed on the UNLV Cosmology Cluster. 
KN is grateful for the hospitality of the Institute for the Physics and Mathematics of the Universe (IPMU), University of Tokyo, where part of this work was done.


\end{document}